\documentclass[11pt]{article}
\usepackage{amsmath}
\usepackage{amsfonts}
\usepackage{epsfig}
\usepackage{amssymb}

\makeatletter \@addtoreset{figure}{section}
\def\thefigure{\thesection.\@arabic\c@figure}
\def\fps@figure{h, t}
\@addtoreset{table}{section}
\def\thetable{\thesection.\@arabic\c@table}
\def\fps@table{h, t}
\@addtoreset{equation}{section}

\makeatother

\newtheorem{thm}{Theorem}[section]
\newtheorem{prop}[thm]{Proposition}

\newfont{\tenbi}{cmbxti10}

\begin{document}
\pagestyle{empty}

\title{Gyroscopic Classical and Quantum
Oscillators interacting with Heat Baths}
\author{
Anthony M. Bloch\thanks{Research partially supported by National
Science Foundation grants DMS 981283 and 0103895 and AFOSR}
\\Department of Mathematics
\\ University of Michigan
\\ Ann Arbor, MI 48109
\\ {\small abloch@math.lsa.umich.edu}
\and Patrick Hagerty\thanks{Research partially supported by
National Science Foundation  and AFOSR.}
\\Department of Mathematics
\\ University of Michigan
\\ Ann Arbor, MI 48109
\\ {\small hagerty@umich.edu}
\and Alberto G. Rojo\thanks{Research partially supported by the
National Science Foundation.}
\\ Dept. of Physics
\\ University of Michigan
\\ Ann Arbor, MI 48109
\\{\small rojoa@umich.edu}
\and Michael I. Weinstein\thanks{Research partially supported by
the National Science Foundation }
\\ Fundamental Mathematics Research Department
\\ Bell Laboratories
\\ 600 Mountain Avenue
\\ Murray Hill, NJ 07974
\\{\small miw@research.bell-labs.com}
}

\maketitle \thispagestyle{empty}

\begin{abstract}
We
 analyze the stability of a gyroscopic oscillator interacting with
a finite- and infinite-dimensional heat bath in both the classical
and quantum cases. We consider a finite gyroscopic oscillator
model
 of a particle in a magnetic field and examine the stability before and after coupling to a heat bath. It
is shown that if the oscillator is gyroscopically stable,
coupling to a sufficiently massive heat bath induces instability.
The meaning of these ideas in the quantum context is discussed.
The model extends the exact diagonalization analysis of an
oscillator and field of Ford, Lewis, and O'Connell to the
gyroscopic setting.

\end{abstract}


\section{Introduction}
In this paper we investigate the stability of a gyroscopically
stabilized system interacting with a finite dimensional heat bath.
More details will be given in a forthcoming paper \cite{BHRWlong}.

A gyroscopically stabilized system is one that is unstable without
gyroscopic forces but becomes stable with the addition of these
forces.
 Infinitesimal dissipative perturbations  are known to induce instability
in Hamiltonian systems that are gyroscopically stabilized;
 see \cite{BKMR}.

Since the origins of dissipation
 ({\it e.g.} friction,
viscosity...) lie in the transfer of energy from one form (energy
 of one subsystem) to another form (that of a second subsystem) of a larger
conservative system, it is natural to expect the analogue of the
above
 destabilization
  phenomenon to be present
within the more fundamental context of conservative systems which
exhibit
 internal energy transfer. In \cite{HBW} and \cite{HBW2},
 we explore this in the context of a
gyroscopic
 oscillating mechanical system coupled to an extended  wave system
 (infinite string). Due to the coupling,
motion within the mechanical system generates waves which can be
  carried off to
infinity. Such {\it radiation damping} has been studied
 in models arising in the theory of quantum resonances, ionization
 type problems and nonlinear waves;
 for more detail see \cite{SofferWeinstein1998b}, \cite{SofferWeinstein1998a}, \cite{SofferWeinstein1999a}, \cite{Kirr},
 and references therein.

As a model of a gyroscopically stabilized system, we consider a
charged  particle stabilized by  a magnetic field. We show that
even for a {\it finite} heat bath sufficient coupling strength
induces instability. This result is quite striking in the sense
that one does not expect a finite heat bath to mimic dissipation.
A graphical criterion is given for determining the onset of instability.
Further we generalize the analysis to the quantum setting. The
infinite limit and oscillator susceptibility are analyzed in the 
forthcoming paper \cite{BHRWlong} mentioned above. 

In the quantum setting we show that while a stable oscillator has
positive energy bound states, a gyroscopically stable oscillator
has both positive and negative energy bound states.
Coupling to the bath with
sufficient coupling strength induces unbound states in the
gyroscopically stable system.
 Applications
of such quantum systems, such as Penning traps, are used to trap
small molecules and obtain extremely precise measurements of
atomic quantities \cite{Geonium}.

Our analysis extends the heat bath analysis
 of \cite{Ford} to the
gyroscopic setting. We show here that there is a beautiful
extension of their graphical (intersection-theoretic) criterion
for stability to a more complex class of curves.

\section{Example of a Chetaev system}
In this section, we discuss a physical realization of the
gyroscopic oscillators discussed in this paper. Consider the
motion of a charged spherical pendulum in a magnetic field whose
linearization is that of a charged planar oscillator in a magnetic
field (for
 details of the full nonlinear system see e.g.
\cite{HBW2}.)


Let ${\bf B}$ be a
divergence-free vector field. Let ${\bf A}$ be the vector
potential, ${\bf B} =\nabla \times {\bf A}.$ Note if we choose
${\bf B}$ to be the constant magnetic field in the direction
normal to the plane of oscillation, the vector potential can be
chosen as , ${\bf A}= \frac{1}{2} {\bf B} \times {\bf q},$ where
${\bf q} =(x,y,0)^T$ is the position of the oscillator.

Assume the oscillator has unit mass and unit charge and that the
speed of light is unity. The Lagrangian, $L:T\mathbb R^2
\rightarrow \mathbb R$, is defined by:

\begin{equation}\label{eq:L}
\begin{split}
L(\bf q, \bf {\dot q}) &= \frac{1}{2} \| {\bf {\dot q}} \|^2 +
{\bf A} \cdot {\bf {\dot q}} - U(\bf q) \\
U(\bf q)&=  \frac{1}{2}(\alpha x^2 + \beta y^2).
\end{split}
\end{equation}
Choosing ${\bf B}$ to be of constant strength $B$, normal to the
plane of oscillation, we obtain the Euler-Lagrange equations of
motion:
\begin{equation}\label{eq:Magnetic field 2d}
\begin{split}
\ddot x -B \dot y + \alpha x &=0\\ \ddot y + B \dot x + \beta y
&=0\,.
\end{split}
\end{equation}

{\bf Remark 1:} If $\alpha$ and $\beta$ are both negative, the oscillator
in a field
system is unstable for small $B$. However, if $B^2 + \alpha +
\beta > 2\sqrt{\alpha \beta},$ the oscillator stabilizes, i.e. the
eigenvalues are on the imaginary axis -- this is what is referred
to as {\it gyroscopic stabilization}. For further details see
below and \cite{BKMR}.

{\bf Remark 2:} If the system is gyroscopically stable it can be
shown that adding a small amount of dissipation to the system
renders it unstable (i.e. there are unstable eigenvalues). For a
more precise statement and generalizations see section
\ref{S:Stability of Chetaev systems}.

\section{Stability of gyroscopic systems}\label{S:Stability of Chetaev systems}
We recall here some general properties of linear systems with
gyroscopic forces. The systems above are examples of such systems.

The general form of a gyroscopic system is
\begin{equation}
M \ddot{\bf q}+ S \dot{\bf q} + V {\bf q}=0,
\label{gengyro}
\end{equation}
where ${\bf q} \in {\mathbb R}^n,$ $M$ is a positive-definite
symmetric $n \times n$ matrix, $S$ is skew, and $V$ is symmetric.
As in \cite{BKMR} we shall call this the {\it Chetaev system} (see
\cite{Chetaev}).

We say the system is {\it gyroscopically stable} if for $S=0$ the
origin is an unstable equilibrium, but for $S\ne 0,$ the origin is
a spectrally stable equilibrium (i.e. the eigenvalues of the
linearized system have non-positive real part). The matrix $S$ is
sometimes referred to as a magnetic term which arises from charged
oscillators in a magnetic field.

An important property of this system is that it is normal form for
a simple mechanical system about a {\it relative equilibrium}
which is given modulo an abelian group. That is, it is the normal
form of a system defined on the cotangent bundle $TQ$ of a
configuration space $Q$ where the Lagrangian is given by kinetic
minus potential energy. One can obtain a similar normal form in
the case of a non-abelian group. (See \cite{BKMR} and
\cite{SimoLewisMarsden1991}.) The magnetic term naturally arises
in the symplectic form when investigating the quotient space.
However, we can obtain the same dynamics from a canonical
symplectic form and an augmented Hamiltonian. This can always be
done by the momentum shifting lemma (see \cite{MarsdenRatiu}).
Classically, the two representations of the Chetaev systems are
equally useful. However, the canonical bracket
 is preferable when quantizing the mechanical system. (See
\cite{BKMR}, \cite{BaillieulLevi}, \cite{Chetaev} for further
physical discussions.)

As for gyroscopic stability, the number of negative eigenvalues of
the quadratic form plays a crucial part as \cite{Chetaev}
discusses and which we summarize in the following proposition:

\begin{prop}
Consider the canonical gyroscopic system $M \ddot {\bf q} + S \dot
{\bf q} + V {\bf q}=0,$ where $M$ is a symmetric positive definite
matrix, $S$ is a skew-symmetric matrix, and $V$ is a symmetric
matrix:
\begin{itemize}
\item {If $V$ has a odd number of negative eigenvalues (counting
multiplicity) then the origin is an unstable equilibrium.}
\item{If $V$ has an even number of negative eigenvalues (counting multiplicity), we can
choose $S$ so that the origin is a spectrally stable equilibrium.}
\end{itemize}
\end{prop}

We omit the proof here.

Gyroscopically stable systems exhibit interesting instability when
perturbed by dissipative forces. Suppose now that $V$ has at least
one negative eigenvalue. A key result of \cite{BKMR} is that
adding small dissipation always yields instability. More precisely
it is shown that

\begin{thm}\label{T:Dissipation}  Under
the above conditions, if we modify the general Chetaev system  by
adding a small Rayleigh dissipation term,
\begin{equation}\label{eq:Dissipation induced instability}
M \ddot{ q} + (S + \epsilon R) \dot{q} + V q = 0
\end{equation}
for small $ \epsilon > 0 $, where $R$ is symmetric and positive
definite, then the perturbed linearized equations  \[ \dot{z} = L
_\epsilon z, \] where $ z = (q, p)$ are spectrally unstable, {\it
i.e.,\/} at least one pair of eigenvalues of $ L _\epsilon $ is in
the right half plane.
\end{thm}

This result builds on basic work of \cite{ThomsonTait},
\cite{Chetaev}, and \cite{Hahn}. We refer to this as {\it
dissipation induced instability}. We see that the Hamiltonian of a
gyroscopically stabilized Chetaev system is indefinite. In this
case, Rayleigh dissipation decreases the value of the Hamiltonian,
but this does not bound the motion of the Chetaev system.  In
particular, the Hamiltonian may decrease to zero, while the
displacement and velocities grow exponentially.

\cite{BKMR} also prove a similar stability result for the
non-abelian case, but the abelian result is sufficient for our
purposes.

\section{Oscillator coupled to a bath}\label{S:oscillator bath}
We now consider the gyroscopic oscillator coupled to a bath of
oscillators via an augmented Lagrangian
\begin{equation}
L_{coupled}=L + L_{bath}
\end{equation}
where $L$ is defined in equation (\ref{eq:L}) and
\begin{eqnarray*}
L_{bath}&=&\sum_{j=1}^N \frac{1}{2} m_j(\dot x_j^2 + \dot y_j^2)\\
&&-\frac{1}{2}  m_j\omega_j^2 ( (x_j -x)^2 + (y_j -y)^2),
\end{eqnarray*}
with  $0<\omega_1< \cdots <\omega_N$ being the characteristic
frequencies of the bath. Similarly, the Hamiltonian can be
augmented in corresponding fashion.

 This model extends the model described by \cite{Ford} to the
gyroscopic setting. As discussed in that paper this model provides
a good physical realization of oscillator-bath coupling. For
simplicity of the computations here we restrict ourselves to the
case $\alpha=\beta$. Similar results hold in the general case. We
also restrict ourselves to the generic situation 
though again similar results hold in
the non-generic setting.

The theorem below in fact generalizes the results of \cite{Ford}
in a rather beautiful fashion, leading to the study of a more
complex intersection problem.

We can show:
\begin{thm}
Consider the case where $\alpha=\beta$ and the equation
\begin{equation}
(\omega\pm \frac{1}{2}B)^2 -(\alpha+\frac{1}{4}B^2) = \sum_{j=1}^N
m_j\frac{\omega_j^2 \omega^2}{\omega^2 -\omega_j^2} \label{eq:
Heat bath theorem}
\end{equation}

Generically we have the following:

The oscillator is stable if there exists $4N+4$ real frequencies
$\omega=\Omega_i\,, i=\pm1,\cdots \pm(2N+2)$ which are solutions
to the equation (\ref{eq: Heat bath theorem}). Let $\Omega_1$ be
the smallest positive solution and let $\Omega_i$ be increasing
with respect to the index $i$.

More precisely we have
\begin{enumerate}
\item[(i)] { In the case of instability of the
oscillator, ($\alpha + \frac{1}{4}B^2 <0$), there are only $4N$
real solutions.  Additionally, there are 2 pairs of imaginary
solutions corresponding to instability.}
\item[(ii)] {In the case of strong stability ($\alpha>0$), there are
$4N+4$ real, normal modes maintaining stability via coupling.  }
\item[(iii)] {  In the case of gyroscopic stability ($\alpha
+\frac{1}{4}B^2>0$), there are two possibilities.  The values of
$m_j$ and $\omega_j$ determine which occurs:

\begin{enumerate}
\item[(a)]{ If $\Omega_1< \omega_1$, then there are $4N+4$
real, normal modes (counting multiplicities) corresponding to
stability.}
\item[(b)]{ If $\Omega_1> \omega_1,$ then there are $4N$ real,
normal modes and $2$ pairs of imaginary mode corresponding
instability in the coupling.  This case occurs for large $m_j$.}
\end{enumerate}}

\end{enumerate}
\end{thm}

{\bf Idea of Proof.} The equations of motion are
\begin{eqnarray*}
\ddot x -B \dot y + \alpha x &=&\sum_{j=1}^N  m_j\omega_j^2 (x_j -x)\\
\ddot y + B \dot x + \beta y &=&\sum_{j=1}^N  m_j\omega_j^2 (y_j -y)\\
\ddot x_j + \omega_j^2 x_j&=& \omega_j^2 x\\
\ddot y_j + \omega_j^2 y_j&=& \omega_j^2 y.
\end{eqnarray*}
We seek normal mode solutions of the form
\begin{eqnarray*}
x(t)= X_0(\omega) e^{i \omega t}; &\quad& y(t)=Y_0(\omega) e^{i \omega t}\\
 x_j(t)= X_j(\omega) e^{i \omega t}; &\quad& y_j(t)=Y_j(\omega) e^{i \omega
t}\,.
\end{eqnarray*}

\begin{figure}
\begin{center}
\epsfig{file=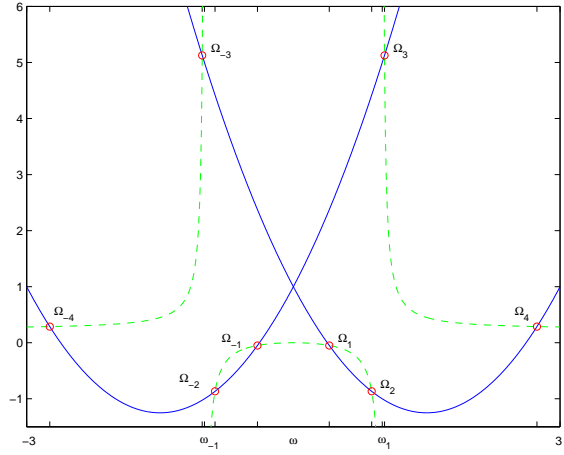,width=7.5cm}
\caption{\label{F:heatbath coupling n=1}{\bf Stable.} $\alpha=-1,$
$B=3$, $m_1=\frac{1}{4},$ and $\omega_1=1.$ }
\end{center}
\end{figure}

In the isotropic case ($\alpha=\beta$), standard normal mode
analysis yields nontrivial solutions  for the characteristic
equation (\ref{eq: Heat bath theorem}). Stability can be insured
if all normal modes are real. The precise results  follow simply
from examining the number of real intersections of the curves on
the left and right sides of equation (\ref{eq: Heat bath
theorem}). For all cases it is helpful to refer to figures
\ref{F:heatbath coupling n=1}, \ref{F:heatbath coupling n=1,
unstable} for the case $N=1$ plotting left hand side and right
hand side of equation (\ref{eq: Heat bath theorem}).

\hfill $\blacksquare$

Details of the proof are given in \cite{BHRWlong}.

\begin{figure}
\begin{center}
\epsfig{file=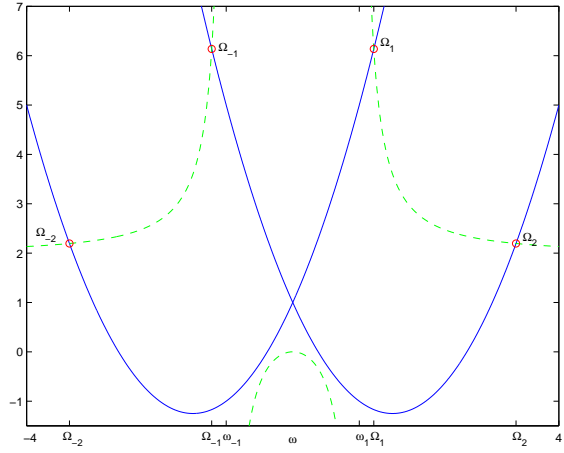,width=7.5cm}
\caption{\label{F:heatbath coupling n=1, unstable}{\bf Unstable.}
$\alpha=-1,$ $B=3$, $m_1=2,$ and $\omega_1=1.$ }
\end{center}
\end{figure}

\section{Gyroscopic quantum oscillators}\label{quantumosc}

The instability results for the classical gyroscopic oscillator
coupled to a heat bath can be extended directly to the quantum
setting. This follows from the general quantization procedure for
a classical system with quadratic Hamiltonian and gyroscopic
terms. We give firstly the general analysis here as this is useful
for understanding how the diagonalization procedure should be
applied in the quantum case.

Consider a general classical Hamiltonian of the form
\begin{equation}
H(q,p)=\frac{1}{2}p^TM^{-1}p+q^TSp+\frac{1}{2}q^TKq
\end{equation}
where $q,p$ are the $n$-dimensional position and momentum vectors
respectively, and $M,S,K$ are constant $n\times n$ matrices with
$M,K$ symmetric and $S$ skew-symmetric. For
\begin{equation}
J=\begin{bmatrix}0&I\\-I&0\end{bmatrix}
\end{equation}
where $I$ is the $n\times n$ identity matrix, the Hamiltonian
equations of motion are given by
\begin{equation}
\frac{d}{dt}\begin{bmatrix}q\\p\end{bmatrix} =J\nabla H=
\begin{bmatrix}-Sq+M^{-1}p\\-Sp-Kq\end{bmatrix}\,.
\end{equation}
Now consider the quantization of the above system. The Heisenberg
equations of motion are given by
\begin{align}
i\hbar\dot{q}&=[q,H]\nonumber\\
i\hbar\dot{p}&=[p,H]
\end{align}
where $q_i,p_i$ obey the standard commutation relations. This
gives
\begin{equation}\label{eq: quantum diagonalization}
i\hbar\frac{d}{dt}\begin{bmatrix}q\\p\end{bmatrix}
=i\begin{bmatrix}-Sq+M^{-1}p\\-Sp-Kq\end{bmatrix}\,.
\end{equation}
just as in the classical setting as an easy computation shows.

The point is that the eigenvalue computation for the equations of
motion in the classical setting is precisely equivalent to that
for the Heisenberg equations of motion. Thus for the oscillator
coupled to the heat bath the numerical computation in the
classical case gives us the correct eigenvalue information for the
quantum case also and we can deduce qualitative behavior as for
the single gyroscopic oscillator in the previous section. Thus,
for sufficiently large coupling, we observe unbound states even
when the uncoupled gyroscopic oscillator exhibits bound states.

One also observes that diagonalizing the Heisenberg equations of
motion is in fact equivalent to diagonalizing the Hamiltonian.

\subsection*{2-D gyroscopic quantum oscillator}
 The quantization of the uncoupled gyroscopic system illustrates
that the stability analysis of the quantum system
  is analogous to the classical
normal mode calculation. However there are some subtleties in the
interpretation of the stability analysis and for this reason we
consider in detail the spectral analysis of an uncoupled quantum
gyroscopic systems. By the analysis of the previous section
similar considerations apply to the full system of the oscillator
coupled to the heat bath.  We follow the Dirac formalism as
described in \cite{Messiah2} for example.

Analogously to the classical coupled oscillators, we define the
quantum Hamiltonian $H$ for an oscillator in a magnetic field of
strength $B$ and isotropic potential,
\begin{eqnarray*}
H&=&\frac{1}{2}(p_x^2 + p_y^2) +\frac{1}{2} (\frac{B^2}{4}+
\alpha)(x^2+y^2)\\
&& + \frac{B}{2} (p_x y - p_y x).
\end{eqnarray*}

For a quantized gyroscopically stable oscillator, we use a
standard change of coordinates to write the Hamiltonian in terms
of normal modes:
\begin{eqnarray*}
A\pm&=&\sqrt{\frac{\omega_0}{4\hbar }}(x\mp iy)+\sqrt{\frac{1}{4
\hbar \omega_0 }}(ip_x\pm p_y)\\
A_\pm^\dagger&=&\sqrt{\frac{\omega_0}{4\hbar }}(x\pm
iy)-\sqrt{\frac{1}{4 \hbar \omega_0 }}(ip_x\mp p_y),
\end{eqnarray*}
where $\omega_0=\sqrt{|\frac{B^2}{4}+ \alpha|}$ and  properties
that $[A_r,A_s]=[A_r^\dagger,A_s^\dagger]=0$ and
$[A_r,A_s^\dagger] =\delta_{rs}$ for $r,s \in \{+,-\}.$  In the
new coordinates, we have
\begin{eqnarray*}
H&=&\hbar (\omega_0 -\frac{B}{2})( A_+^\dagger A_+ +\frac{1}{2}) +\\
&&\hbar(\omega_0 +\frac{ B}{2}) (A_-^\dagger A_-+\frac{1}{2})\,.
\end{eqnarray*}
We verify that the new frequencies $\omega_0 \pm \frac{B}{2}$ are
exactly the classical frequencies from the normal mode
calculation. In the Heisenberg picture, we have
\begin{eqnarray*}
i \hbar \dot A^\dagger_\pm &=& [A^\dagger_{\pm},H]=-\hbar(\omega_0
\mp \frac{B}{2} ) A^\dagger_\pm \\
A^\dagger_\pm &=& e^{i (\omega_0 \mp \frac{B}{2} ) t}
A^\dagger_\pm (t=0).
\end{eqnarray*}
 The eigenvalues of the
Hamiltonian, $E_{n_1,n_2}$ are
\begin{equation*}
E_{n_1,n_2} = \hbar \omega_0(n_1 + n_2+1) +\hbar \frac{B}{2}
(n_1-n_2),
\end{equation*}
where $n_1,n_2 \in \{0,1,2,\cdots\}.$ In the case of gyroscopic
stability, we have $\frac{B}{2} > \omega_0,$ and the energy
spectrum is indefinite and unbounded above and below. In the case
of pure stability ($\alpha>0$), we have $\frac{B}{2} < \omega_0,$
and the energy spectrum in positive definite.

In the case of a quantized unstable gyroscopic oscillator,
($\frac{B}{2} +\alpha <0$), it can be shown there are no bound
eigenstates.

Similarly in the case of a quantum oscillator coupled to a finite
heat bath we can use the change of variables in equation (\ref{eq:
quantum diagonalization}) to compute the characteristic
frequencies, thus reducing the coupled problem to a system of
independent quantum oscillators. Thus the spectrum of the
Hamiltonian of the coupled quantum system is positive if and only
if the classical system is strongly stable; the spectrum is
discrete with both positive and negative eigenvalues if and only
if the classical system is gyroscopically stable; and finally, we
have unbounded eigenfunctions if and only if the classical system
in unstable.

\section{Conclusions}
Often one models dissipation or friction by coupling a mechanical
system to a thermal reservoir.  We have modeled the thermal
reservoir as a collection of oscillators coupled to the mechanical
system.  If the reservoir is sufficiently massive, then the
coupling may induce instability in the mechanical system.
Equivalently, if the reservoir contains sufficiently low
frequencies, then similar instabilities may arise.

Unlike other models of dissipation (i.e. Rayleigh), the
conservative nature of the thermal reservoir allows us to
naturally extend our results to the quantum setting.  We interpret
stability in the quantum setting as corresponding to bound
eigenstates.  A gyroscopically, stabilized, quantum Chetaev system
exhibits this stability while having an unbounded spectrum.

\bibliographystyle{alpha} 
\bibliography{citations} 

\end{document}